\documentclass[1p,times,letterpaper,aas_macros]{elsarticle}
\usepackage{lipsum}
\usepackage{amsmath}
\usepackage{mathtools}
\usepackage{wrapfig}
\usepackage{textcomp}
\usepackage[numbers]{natbib}
\RequirePackage[colorlinks,citecolor=blue,urlcolor=blue,linkcolor=blue]{hyperref}
\RequirePackage{url}
\usepackage{etoolbox}
\usepackage[table]{xcolor}
\hypersetup{linkbordercolor=green}
\usepackage{caption}
\usepackage{adjustbox}
\usepackage{upgreek}
\usepackage{pifont}% http://ctan.org/pkg/pifont
\usepackage{color,soul}
\usepackage{float}

\floatstyle{plaintop}
\restylefloat{table}

\usepackage[bottom]{footmisc}
\usepackage{enumitem}
\usepackage{lineno}
\graphicspath{{Figures/}}

\journal{Astroparticle Physics}

\title{Cosmic antihelium-3 nuclei sensitivity of the GAPS experiment}

\author[CU]{N.~Saffold\corref{cor1}}
\ead{nas2173@columbia.edu}
\author[Northeastern,SLAC]{T.~Aramaki}
\author[UCLA]{R.~Bird}
\author[INFN-Trieste,IFPU]{M.~Boezio}
\author[UCSD]{S.E.~Boggs}
\author[INFN-Trieste]{V.~Bonvicini}
\author[INFN-Napoli]{D.~Campana}
\author[SSL]{W.W.~Craig}
\author[UHM]{P.~von~Doetinchem}
\author[UCLA]{E.~Everson}
\author[ORNL]{L.~Fabris}
\author[ISAS-JAXA]{H.~Fuke}
\author[CU]{F.~Gahbauer}
\author[UCLA]{I.~Garcia}
\author[UHM]{C.~Gerrity}
\author[CU]{C.J.~Hailey}
\author[UCLA]{T.~Hayashi}
\author[Shinshu]{C.~Kato}
\author[Tokai]{A.~Kawachi}
\author[Tokai]{S.~Kobayashi}
\author[ISAS-JAXA]{M.~Kozai}
\author[INFN-Trieste,UTrieste]{A.~Lenni}
\author[SSL]{A.~Lowell}
\author[INFN-Pavia,UBergamo]{M.~Manghisoni}
\author[INFN-Rome,URoma]{N.~Marcelli}
\author[PSU]{S.I.~Mognet}
\author[Shinshu]{K.~Munakata}
\author[INFN-Trieste,IFPU]{R.~Munini}
\author[AoyamaU]{Y.~Nakagami}
\author[Helio]{J.~Olson}
\author[UCLA]{R.A.~Ong}
\author[INFN-Napoli]{G.~Osteria}
\author[MIT]{K.~Perez}
\author[CU]{I.~Pope}
\author[UCLA]{S.~Quinn}
\author[INFN-Pavia,UBergamo]{V.~Re}
\author[CU]{M.~Reed}
\author[INFN-Pavia,UBergamo]{E.~Riceputi}
\author[MIT]{B.~Roach}
\author[MIT]{F.~Rogers}
\author[UCLA]{J.L.~Ryan}
\author[INFN-Napoli,UNapoli]{V.~Scotti}
\author[KanagawaU]{Y.~Shimizu}
\author[INFN-Pavia,UBergamo]{M.~Sonzogni}
\author[INFN-Rome,URoma]{R.~Sparvoli}
\author[UHM]{A.~Stoessl}
\author[INFN-Firenze]{A.~Tiberio}
\author[INFN-Firenze]{E.~Vannuccini}
\author[AoyamaU]{T.~Wada}
\author[MIT]{M.~Xiao}
\author[ISAS-JAXA]{M.~Yamatani}
\author[AoyamaU]{A.~Yoshida}
\author[ISAS-JAXA]{T.~Yoshida}
\author[INFN-Trieste]{G.~Zampa}
\author[UCLA]{J.~Zweerink}

\cortext[cor1]{Corresponding author: nas2173@columbia.edu}

\address[CU]{Columbia Astrophysics Laboratory, Columbia University, 550 W 120th St, New York, NY 10027, USA}
\address[Northeastern]{Northeastern University, Boston, MA 02115, USA}
\address[SLAC]{Stanford Linear Accelerator Center, 2575 Sand Hill Rd, Menlo Park, CA 94025, USA}
\address[UCLA]{Department of Physics and Astronomy, University of California, Los Angeles, CA 90095, USA}
\address[INFN-Trieste]{INFN, Sezione di Trieste, I-34149 Trieste, Italy}
\address[IFPU]{IFPU, I-34014 Trieste, Italy}
\address[UCSD]{University of California, San Diego, La Jolla, CA 90037, USA}
\address[INFN-Napoli]{INFN, Sezione di Napoli, I-80126 Naples, Italy}
\address[SSL]{Space Sciences Laboratory, University of California, Berkeley, 7 Gauss Way, Berkeley, CA 94720, USA}
\address[UHM]{Department of Physics and Astronomy, University of Hawaii at Manoa, 2505 Correa Rd, Honolulu, HI 96822}
\address[ORNL]{Oak Ridge National Laboratory, Oak Ridge, TN 37831, USA}
\address[ISAS-JAXA]{Institute of Space and Astronautical Science, Japan Aerospace Exploration Agency (ISAS/JAXA), Sagamihara, Kanagawa 252-5210, Japan}
\address[Shinshu]{Shinshu University, Matsumoto, Nagano 390-8621, Japan}
\address[Tokai]{Tokai University, Hiratsuka, Kanagawa 259-1292, Japan}
\address[UTrieste]{Universit\`{a} di Trieste, I-34127 Trieste, Italy}
\address[INFN-Pavia]{INFN, Sezione di Pavia, I-27100 Pavia, Italy}
\address[UBergamo]{Universit\`{a} di Bergamo, I-24044 Dalmine (BG), Italy}
\address[INFN-Rome]{INFN, Sezione di Rome ``Tor Vergata'', I-00133 Rome, Italy}
\address[URoma]{Universit\`{a} di Roma ``Tor Vergata'', I-00133 Rome, Italy}
\address[PSU]{Pennsylvania State University, University Park, PA 16802 USA}
\address[AoyamaU]{Aoyama Gakuin University, Sagamihara, Kanagawa 252-5258, Japan}
\address[Helio]{Heliospace Corporation, Berkeley, CA 94710, USA}
\address[MIT]{Department of Physics, Massachusetts Institute of Technology, 77 Massachusetts Ave, Cambridge, MA 02139, USA}
\address[UNapoli]{Universit\`{a} di Napoli Federico II, I-80138 Naples, Italy}
\address[KanagawaU]{Kanagawa University, Yokohama, Kanagawa 221-8686, Japan}
\address[INFN-Firenze]{INFN, Sezione di Firenze, I-50019 Sesto Fiorentino, Florence, Italy}

\begin{document}
\begin{abstract}
The General Antiparticle Spectrometer (GAPS) is an Antarctic balloon experiment designed for low-energy (0.1--0.3\,GeV/$n$) cosmic antinuclei as signatures of dark matter annihilation or decay. GAPS is optimized to detect low-energy antideuterons, as well as to provide unprecedented sensitivity to low-energy antiprotons and antihelium nuclei. The novel GAPS antiparticle detection technique, based on the formation, decay, and annihilation of exotic atoms, provides greater identification power for these low-energy antinuclei than previous magnetic spectro\-meter experiments. This work reports the sensitivity of GAPS to detect antihelium-3 nuclei, based on full instrument simulation, event reconstruction, and realistic atmospheric influence simulations. The report of antihelium nuclei candidate events by AMS-02 has generated considerable interest in antihelium nuclei as probes of dark matter and other beyond the Standard Model theories. GAPS is in a unique position to detect or set upper limits on the cosmic antihelium nuclei flux in an energy range that is essentially free of astrophysical background. In three 35-day long-duration balloon flights, GAPS will be sensitive to an antihelium flux on the level of $1.3^{+4.5}_{-1.2}\cdot 10^{-6}$\,m\textsuperscript{-2}sr\textsuperscript{-1}s\textsuperscript{-1}(GeV/$n$)\textsuperscript{-1} (95\% confidence level) in the energy range of 0.11--0.3\,GeV/$n$, opening a new window on rare cosmic physics.
\end{abstract}
\maketitle

\nolinenumbers

\section{Introduction}
Astrophysical observations indicate that dark matter is about five times more abundant than baryonic matter~\cite{Planck2020}, but the fundamental nature of dark matter has not been uncovered. Cosmic antinuclei are excellent probes for dark matter models that annihilate or decay in the Galactic halo, including many models that evade detection in collider, direct, or other indirect searches~\cite{Doetinchem_2020}. AMS-02 has announced the observation of several high-momenta ($>$10\,GeV/$c$) candidate an\-tihelium-3 and antihelium-4 nuclei events~\cite{Ting2016}. Data taking, analyses, and interpretation of these events are still ongoing. GAPS is an indirect dark matter detection experiment optimized to detect low-energy (0.1--0.3\,GeV/$n$) cosmic antiprotons, antideuterons, and antihelium using a series of Antarctic long-duration balloon (LDB) flights. GAPS will be complementary to AMS-02, as it has orthogonal systematic uncertainties and operates in the crucial lower-energy range where the predicted contribution from new-physics models compared to astrophysical background is highest. One key advantage of the Antarctic flight path of GAPS, which BESS-Polar similarly benefited from, is the low geomagnetic cutoff compared to the trajectory of AMS-02 on the International Space Station. At least three GAPS LDB flights are planned, with the first launch date anticipated for December 2022.

The flux of antinuclei due to dark matter annihilation and decay can be estimated based on dark matter density profiles in the Galaxy, dark matter annihilation and decay channels, and hadronization, coalescence and Galactic propagation models. Over the past few decades, the cosmic antiproton spectrum has been measured by experiments such as BESS~\cite{Yamamato_2013,Abe_2012_Pbars}, CAPRICE98~\cite{Bergstr_m_2000}, PAMELA~\cite{PAMELA_2010,Adriani_2013}, and AMS-02~\cite{AMS_Aguilar_2016}. A possible excess in the AMS-02 antiproton spectrum could be consistent with 20--80\,GeV dark matter, but the significance of these analyses depend on interpretation of theoretical and experimental systematic uncertainties~\cite{Cui_2017,Cuoco_2017,Reinert_2018,Cuoco_2019,Cholis_2019,Boudaud_2020}. At low-energies, the production of secondary antinuclei from cosmic-ray interactions with the interstellar medium is kinematically suppressed. A variety of dark matter models predict antideuteron fluxes~\cite{Donato2000,Baer_2005,Donato_2008,Ibarra_2013} orders of magnitude above the astrophysical background in the energy range below approximately~1\,GeV/$n$. Naively, any model explaining the AMS-02 antihelium nuclei candidate events would overproduce both antiprotons and antideuterons. Prior to AMS-02, BESS-Polar set an exclusion limit on the antihelium to helium flux ratio of $1.0\cdot 10^{-7}$ in the range of 1.6--14\,GV~\cite{Abe_2012}, the most stringent upper limit on the antihelium flux prior to the tantalizing AMS-02 reports. Attempts to explain the AMS-02 antihelium candidate events predict fluxes in the GAPS energy range covering many orders of magnitude. These range from standard cosmic rays with heavily-tuned formation models~\cite{Blum_2017,Tomassetti_2017}, to new DM annihilation channels~\cite{Cirelli_2014, Carlson_2014, Coogan_2017, Korsmeier_2018, Ding_2019, winkler2020dark}, or even the existence of an antistar within our Galaxy~\cite{2018arXiv180808961P,Khlopov2000}. The GAPS antihelium measurement will provide crucial information to constrain these models, and the detection of low-energy antihelium nuclei would be an exciting sign of new physics.

This work presents the projected sensitivity of GAPS to detect cosmic antihelium-3 nuclei, and will serve as the foundation for future GAPS antihelium-4 nuclei sensitivity studies. An overview of the GAPS experiment is outlined in Sec.~\ref{s-gaps}. The simulation framework is described in Sec.~\ref{s-sim}. The instrument simulations and identification technique are presented in Sec.~\ref{s-id}. The atmospheric simulations to estimate background fluxes and the antihelium-3 nuclei sensitivity of GAPS using three LDB flights are presented in Sec.~\ref{s-sens}. Conclusions and the outlook for the GAPS experiment are discussed in Sec.~\ref{s-conc}.

\section{The GAPS experiment\label{s-gaps}}
\subsection{Instrument overview}
\begin{figure}
\centering
\includegraphics[width=\linewidth]{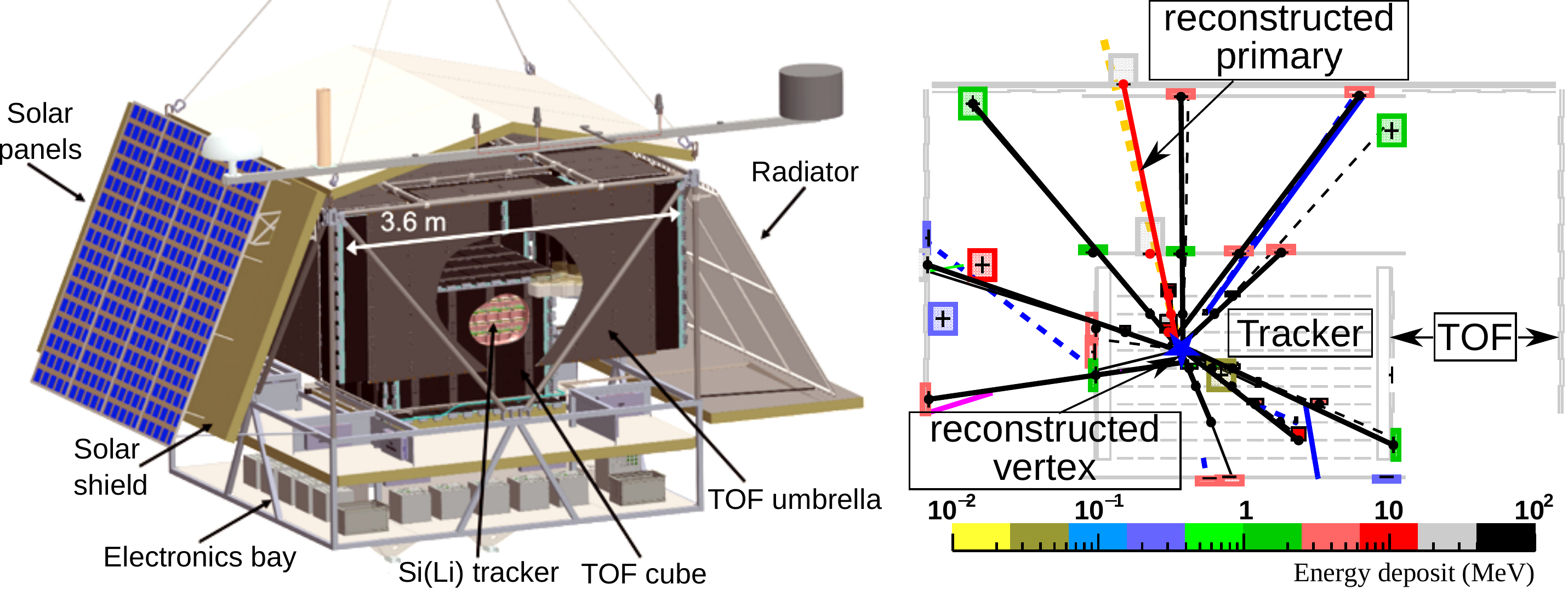}
\caption{\label{f-payloadandrecoevent}\textit{Left:}~Schematic overview of the GAPS instrument: two layers of plastic scintillator form the inner TOF ``cube'' and outer TOF ``umbrella''. The inner TOF cube encapsulates a tracker composed of 10~layers of 1000~Si(Li) detectors. \textit{Right:} Antihelium-3 nucleus event topology: the orange dashed and red solid line indicate the simulated and the reconstructed primary antihelium-3 nuclei, respectively. The blue star designates the reconstructed stopping vertex. The thick solid black lines demonstrate the reconstructed tracks emerging from the stopping vertex. Thin black solid (dashed) lines represent secondary $\pi^+$ ($\pi^-$), blue solid (dashed) lines represent secondary positrons (electrons), and magenta solid (dashed) lines represent secondary $\mu^+$ ($\mu^-)$ from the simulation. The colored boxes show the energy depositions of the registered hits. The color of the box indicates the amount of energy deposited, and the size of the boxes correspond to the estimated error in position.}
\end{figure}

The GAPS experiment is designed to detect cosmic antinuclei during a series of LDB flights at high-altitude (${\approx}37$\,km) above Antarctica. The GAPS instrument consists of a particle tracker surrounded by a time-of-flight (TOF) system (Fig.~\ref{f-payloadandrecoevent}). The TOF system consists of 196 plastic scintillator paddles arranged into an outer ``umbrella'' and an inner ``cube'' separated by a minimum distance of 0.95\,m. Each plastic scintillator paddle is 6.35\,mm thick and 16\,cm wide. The umbrella consists of 1.8\,m length paddles, whereas the cube uses 1.8\,m, 1.56\,m, and 1.1\,m lengths. The prototype TOF paddles have demonstrated a time-of-flight resolution of better than 400\,ps, and high-speed trigger and veto capabilities~\cite{quinn2020}. The inner TOF cube encapsulates the particle tracker formed from 1000 10\,cm diameter, 2.5\,mm thick lithium-drifted silicon (Si(Li)) detectors, arranged into ten tracking planes. Each Si(Li) detector has a cylindrical geometry with an active area of about 70\,cm\textsuperscript{2} that is segmented into eight single-sided strips of equal area~\cite{KP2018,Kozai,Field,SaffoldPassivation}. An oscillating heat pipe system~\cite{Okazaki2018} in conjunction with a rotator to keep the radiator pointed away from the sun is used to cool the Si(Li) detectors to the requisite operational temperature ({$\approx$}--40\,\textdegree{C}). GAPS has a large instrumental acceptance -- which is necessary for rare signal searches -- and provides multiple identification techniques to reject cosmic-ray backgrounds.

\subsection{GAPS Identification technique}
GAPS uses a novel detection technique based on the formation, de-excitation, and annihilation of exotic atoms to identify cosmic antinuclei~\cite{Mori_2002,Hailey2009}. Protons, $\upalpha$-particles, and antiprotons are the main backgrounds for the antihelium-3 nuclei search because protons and $\upalpha$-particles are the most abundant cosmic-ray species and antiprotons are the most abundant antinuclei species. Higher-charge particles are reliably rejected by the trigger algorithm. The GAPS antihelium nuclei identification scheme relies on reconstructing the antihelium nuclei's ionization losses before the annihilation to reject antiprotons and protons, as well as the multiplicity, velocity, and angular distribution of particles emerging from the stopping vertex (secondaries) to reject $\upalpha$-particles and protons.

The right panel of Fig.~\ref{f-payloadandrecoevent} shows a typical antihelium-3 nucleus event topology inside the GAPS instrument. The incident primary particle traverses the TOF system, which triggers the instrument readout. The TOF also timestamps the energy depositions in the scintillator paddles, which enables the measurement of the primary particle's velocity $\beta$. Afterward, the particle continues into the Si(Li) tracker, where it slows down through ionization losses ($\text d E/\text d x$) in the tracker material. The Si(Li) detectors measure the energy depositions on the track, which increase as approximately $Z^{2}/\beta^2$ as the particle slows, where $Z$ refers to the primary particle's charge. The charge-dependence of the ionization losses is a crucial component of the GAPS antihelium nuclei detection concept. For the same $\beta$, antihelium nuclei will deposit four times as much energy as antiprotons and protons (Sec.~\ref{sec-eventvariables}).
 
\begin{figure}
\centering
\includegraphics[width=0.5\linewidth]{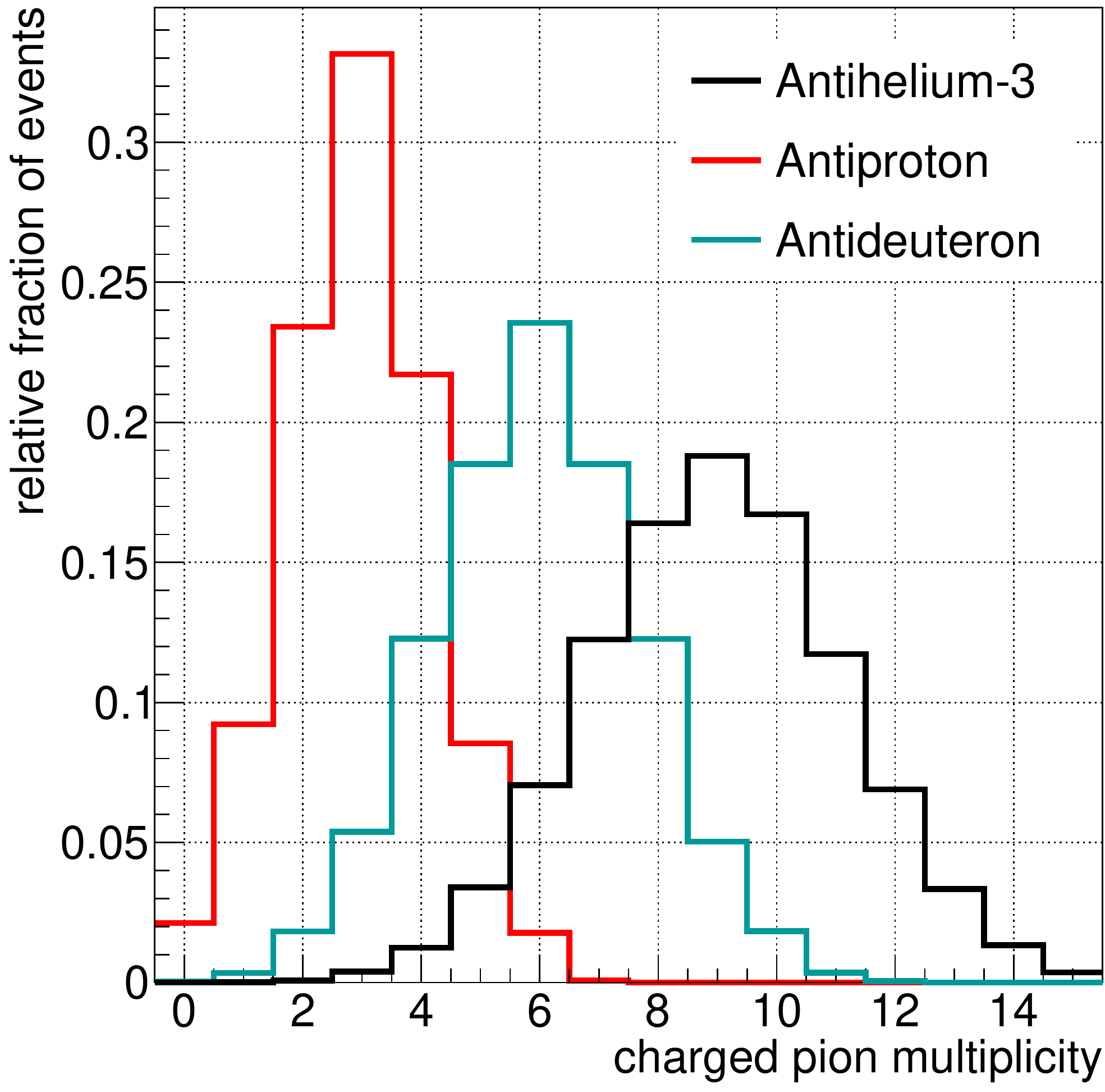}
\caption{\label{sec-pionmultiplicity}Simulated charged-pion multiplicity produced in annihilations of antiprotons, antideuterons, and antihelium-3 nuclei at rest in silicon using {\tt Geant4v10\hspace{-0.12em}.\hspace{-0.12em}6\hspace{-0.12em}.\hspace{-0.12em}p02}.}
\end{figure}

When the kinetic energy of an antinucleus is comparable to the binding energy of a target atom, the antinucleus is captured by the target material with near-unity probability, forming an exotic atom in a highly excited state. Within $\mathcal{O}$(1\,ns), the exotic atom de-excites emitting Auger electrons and X-rays before the antinucleus annihilates with the target material, producing pions and protons. The lower-energy radiative transitions are in the 20--100~keV range and can be detected with the Si(Li) detectors~\cite{Aramaki2013}. The charged-particle annihilation products (primarily pions) can be tracked with both the tracker and the TOF, enabling energy deposition and velocity measurements. The number of pions emerging from the stopping vertex scales with the number of antinucleons in the annihilating antinucleus, which helps distinguish antinuclei event signatures (Fig.~\ref{sec-pionmultiplicity}). 

\section{Simulation \label{s-sim}}
A dedicated {\tt Geant4}-based simulation~\cite{Agostinelli2003250,2006ITNS...53..270A} and analysis framework was developed to model the GAPS payload and its interactions with cosmic-rays. The simulated geometry includes active detector components (Si(Li) detectors in the tracker, plastic scintillator paddles in the TOF) and their electronics response, and the passive structural components. The simulation assumes a time-of-flight resolution of 300\,ps -- the target timing resolution for the GAPS TOF system. The identification analysis is insensitive to the TOF resolution to the extent that using the demonstrated TOF resolution does not change the result. To simulate the physics processes, the {\tt FTFP\textunderscore BERT\textunderscore HP} physics list was used in {\tt Geant4v10\hspace{-0.12em}.\hspace{-0.12em}6\hspace{-0.12em}.\hspace{-0.12em}p02}. The implemented {\tt Geant4} physics processes for antiproton annihilations at-rest, as described by the Fritiof model~\cite{Galoyan_2015}, were validated by comparison to available data from accelerator-based experiments~\cite{AmslerCrystal}.

Primary antihelium-3 nuclei and their dominant backgrounds (antiprotons, protons, and $\upalpha$-particles) were generated from the top of the instrument (TOI). The simulated primary particles were generated with a uniform velocity distribution ($0.1 < \beta_{\text{TOI,gen}} < 1$) and an isotropic angular distribution from the surface of a 4.4\,m side-length cube encapsulating the GAPS instrument. The geometrical acceptance of the instrument is calculated following the standard approach from~\cite{Sullivan_1971}. For this study, $10^{11}$ protons, $6 \cdot 10^{9}$ $\upalpha$-particles, $5 \cdot 10^{8}$ antiprotons, and $2 \cdot 10^{8}$ antihelium-3 nuclei were generated. The simulations made use of the intended GAPS trigger scheme that is designed to reject high-velocity and high-charge particles. Identification studies focused on rejecting the dominant $|Z|=1$ and $|Z|=2$ backgrounds, since the trigger and preselection (Sec.~\ref{sec-preselection}) criteria reliably reject background contamination from heavier nuclei, such as carbon and boron. The trigger scheme requires that the TOF energy depositions are in the range of slow-moving $|Z|=1$ or $|Z|=2$ to reject minimally ionizing (high-velocity) and high-charge particles. In addition, at least eight hits in the combined TOF system, with at least three hits each in the TOF umbrella and TOF cube, are required to focus the data taking on annihilating antinuclei and reject non-annihilating positively charged particles with low secondary multiplicity. The trigger algorithm provides a rejection factor of approximately 700 and 50 for protons and $\upalpha$-particles, respectively, while retaining a significant fraction (${>}50\%$) of incoming antinuclei~\cite{quinn2020}.

\section{Particle identification\label{s-id}}

\subsection{Event reconstruction}
\label{sec-eventreco}
As illustrated in the right panel of~Fig.~\ref{f-payloadandrecoevent}, the reconstruction algorithm reconstructs the primary particle's trajectory, its stopping vertex, and secondary tracks emerging from the vertex. The GAPS algorithm starts by identifying the earliest hits in the TOF and then iteratively adds hits from the tracker that are spatially and energetically consistent with the primary track. In the next step, a search for the annihilation star signature is performed along the primary particle trajectory to identify secondary tracks emerging from the stopping vertex. In the final step, a mini\-mization procedure is performed to find the most likely stopping vertex. This algorithm has an efficiency of $>$80\% to identify antinuclei that stopped inside the tracker volume and reconstructs the stopping vertex to within 60\,mm for about 70\% of the events. The identification analysis (Sec.~\ref{sec-eventvariables}) takes into account the typical energy loss in the instrument to determine the primary particle's reconstructed velocity at TOI, $\beta_{\text{TOI,rec}}$. The combination of the reconstruction algorithm with the energy loss correction achieves a velocity resolution of less than 5\% in the relevant velocity range ($0.1 < \beta_{\text{TOI,gen}} < 0.6$).

\subsection{Analysis preselection}
\label{sec-preselection}
Before the identification analysis, preselection criteria (or cuts) are applied to select well-reconstructed events. These preselection cuts require at least one hit from the reconstructed primary track in each of the TOF umbrella and TOF cube, a reconstructed stopping vertex in the tracker, no more than one active volume on the reconstructed primary track without a registered hit, and the energy depositions on the primary track consistent with a $|Z|=1$ or $|Z|=2$ particle at the reconstructed velocity. This last cut is essential to suppress events, typically of low velocity, where the primary particle annihilates in the TOF.

\subsection{Identification analysis}
\label{sec-eventvariables}

\begin{figure}
\centering
\begin{minipage}{0.48\textwidth}
\centering
\includegraphics[width=1\linewidth]{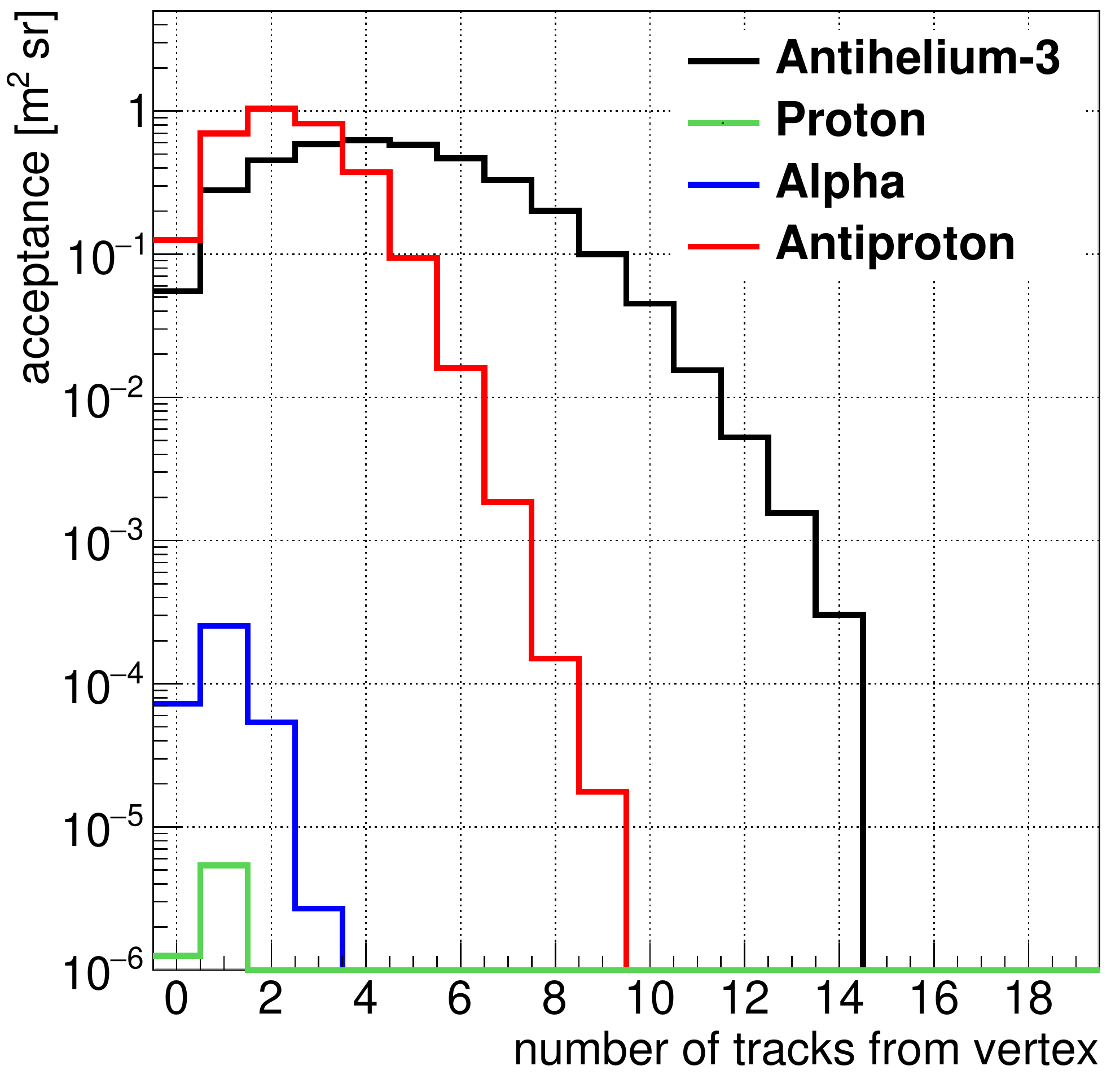}
\end{minipage}
\begin{minipage}{.48\textwidth}
\centering
\includegraphics[width=1\linewidth]{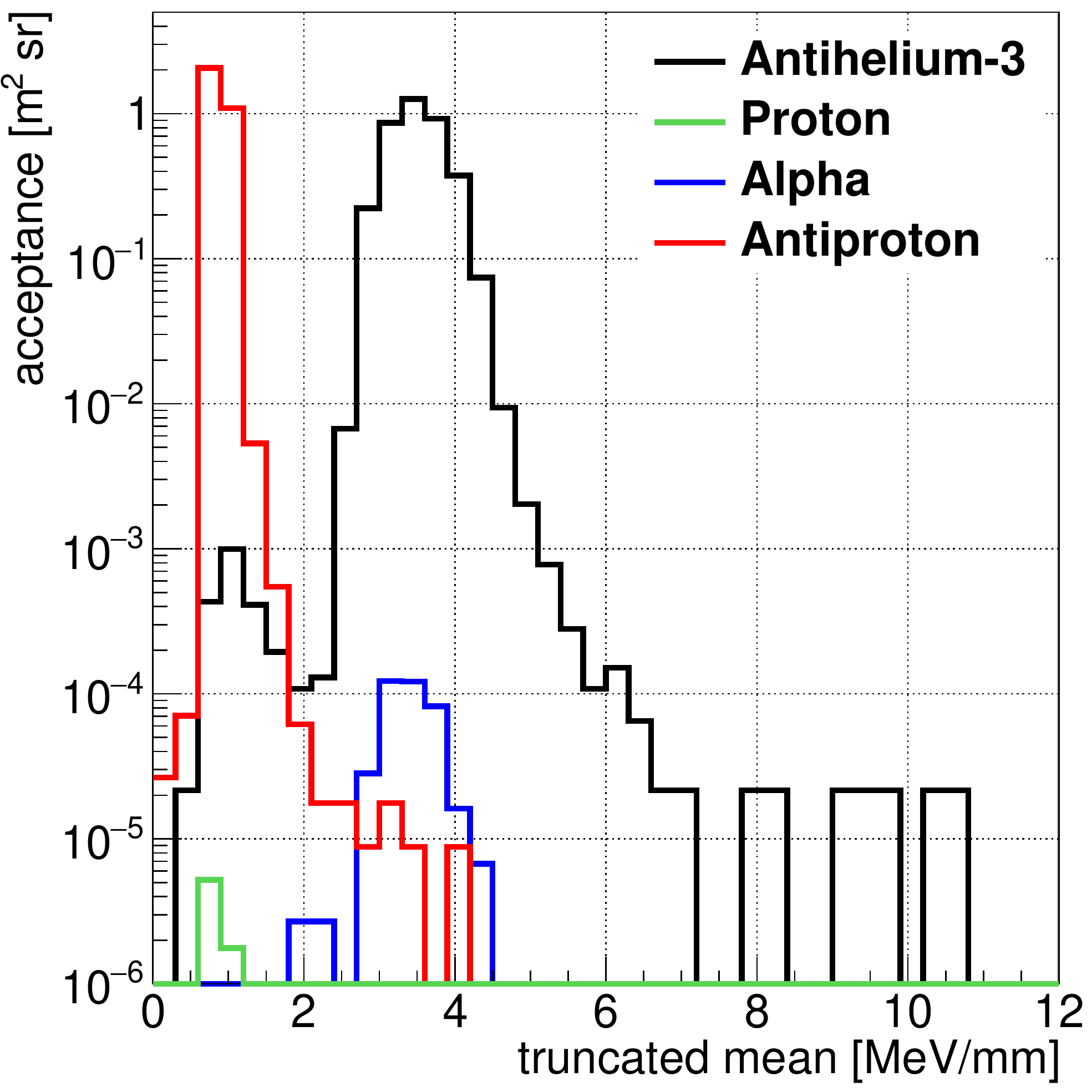}
\end{minipage}
\caption{\label{f-rejpower}Examples of the estimated acceptance for two identification variables for antihelium-3 nuclei, antiprotons, protons, and $\upalpha$-particles passing trigger and quality cuts (velocity range $0.39 < \beta_{\text{TOI,gen}} < 0.41$): \textit{Left:} Number of reconstructed tracks from the stopping vertex. \textit{Right:} Truncated mean energy deposition on the primary track.}
\end{figure}

Using the information from the event reconstruction (Sec.~\ref{sec-eventreco}), the following seven variables are combined in an identification analysis to identify antihelium-3 nuclei: 

\begin{description}
\item [Truncated mean energy deposition:] this variable is calculated by sorting the primary particle's energy depositions normalized to the pathlength in the corresponding volume ($\text d E/\text d x$) in ascending order, followed by averaging the lower half of the $\text d E/\text d x$ values. Since $\text d E/\text d x$ scales with $Z^2/\beta^2$, selecting the primary hits with lower $\text d E/\text d x$ values ensures that energy depositions close to the stopping vertex, when the particle's $\beta$ has decreased from ionization losses, are not included in the $\text d E/\text d x$ calculation. This allows determining the primary particle's charge $|Z|$ when combining it with the reconstructed primary particle's velocity $\beta_{\text{TOI,rec}}$ (Fig.~\ref{f-rejpower}, right). It provides major rejection power to distinguish antihelium-3 nuclei events from $|Z|=1$ particles (antiprotons and protons) as well as carbon and other heavier nuclei.
\item [Number of secondary tracks from the vertex:] this is determined using the multiplicity of reconstructed secondary tracks emerging from the vertex. For positively charged particles that pass the trigger, the secondary multiplicity is much lower than for antihelium-3 nuclei (Fig.~\ref{f-rejpower}, left).
\item [Total number of hits:] this variable is determined by counting the total number of registered hits in the TOF and tracker. Similar to the number of secondary tracks from the vertex, the total number of hits provides means to assess the secondary multiplicity. The number of hits for antihelium-3 nuclei annihilation events is much higher than for positively charged particles.
\item [Total energy deposition:] this variable is calculated by summing all energy depositions in the TOF and tracker during an event. Antinuclei that annihilate in the tracker deposit a large amount of energy in the instrument.
\item [Average velocity of secondary tracks:] the velocity of each secondary track emerging from the vertex is determined using the timing information associated with the reconstructed stopping vertex and the successive hits in the TOF. The average velocity of the secondary tracks is higher for antinuclei than for nuclei because the annihilation process enables the formation of relativistic pions, whereas inelastic collisions of particles must conserve baryon number and are more likely to produce lower-velocity protons.
\item [Isotropy of secondary tracks:] this variable is determined by averaging the cosine of the angle between the primary particle's direction and the direction reconstructed between individual tracker hits and the stopping vertex. Antinuclei annihilation at-rest has a more isotropic secondary signature while inelastic collisions of high-velocity protons and $\upalpha$-particles are more forward-boosted.
\item [Primary column density:] this variable evaluates the grammage traversed by the primary particles from the top of the instrument to the stopping vertex. For the same primary velocity, antihelium-3 nuclei will typically traverse 25\% less grammage than antiprotons, protons, and $\upalpha$-particles before stopping.
\end{description}

Next, probability distributions of these variables for the different particle types are created as a function of the primary particle's generated velocity $\beta_{\text{TOI,gen}}$ and the cosine of the generated zenith angle $\cos(\theta_{\text{TOI,gen}})$. The construction of these probability distributions accounts for velocity resolution effects by introducing a Gaussian smearing of the probability distributions, which depends on the $\beta$ resolution as a function of the primary's velocity. The probability distributions are used to perform an identification analysis to determine the likelihood of each event being a signal event relative to a background event. The likelihood function $\mathcal{P}^{a}$ describing the likelihood for a given event to be a particular particle species $a$ is calculated as:
\begin{equation}
\mathcal{P}^{a}=\sqrt[N]{\prod_{i}^N P^{a}_{i}\left(\beta_{\text{TOI,rec}},\cos(\theta_{\text{TOI,rec}})\right)}.
\end{equation}
Here, $P^{a}_{i}(\beta_{\text{TOI,rec}},\cos(\theta_{\text{TOI,rec}}))$ is a probability distribution for one of the $N=7$ identification variables, indexed by $i$ and evaluated at an event's $\beta_{\text{TOI,rec}}$ and $\cos(\theta_{\text{TOI,rec}})$ for a certain particle species $a$. These $\mathcal{P}^{a}$ values are used to construct the likelihood ratio $L$:
\begin{equation}
\label{eqn:llh-ratio}
L = \frac{\mathcal{P}^{{}^3\overline{\text{He}}}}{\mathcal{P}^{{}^3\overline{\text{He}}}+\mathcal{P}^{\overline{p}}+\mathcal{P}^{p}+\mathcal{P}^{\alpha}}
\end{equation}

In the analysis that follows, the natural logarithm of the ratio $-\ln(L)$ is used as the identification variable. A low $-\ln(L)$ value indicates a high probability of being an antihelium-3 nucleus event. Before evaluating the likelihood ratio, two additional cuts were applied. Candidate antihelium-3 nucleus events are required to have a truncated mean energy deposition (Fig.~\ref{f-rejpower}, right) corresponding to a charge of $|Z|=2$, to ensure an unambiguous charge measurement of the primary. Furthermore, candidate events are required to have a reconstructed velocity $\beta_{\text{TOI,rec}}$ in the range of 0.3--0.6 to assure that a candidate antihelium-3 nucleus could stop inside the TOF cube. This analysis was conducted for three different $\cos(\theta_{\text{TOI,rec}})$ ranges ($\cos(\theta_{\text{TOI,rec}})=[0,1/3], $ $[1/3,2/3], [2/3,1]$). For each angular range, cuts on $-\ln(L)$ were optimized to reject background events while maximizing GAPS's antihelium-3 nuclei acceptance (Sec.~\ref{sec-atmosphere}).

\begin{figure}
\centering
\includegraphics[width=0.5\linewidth]{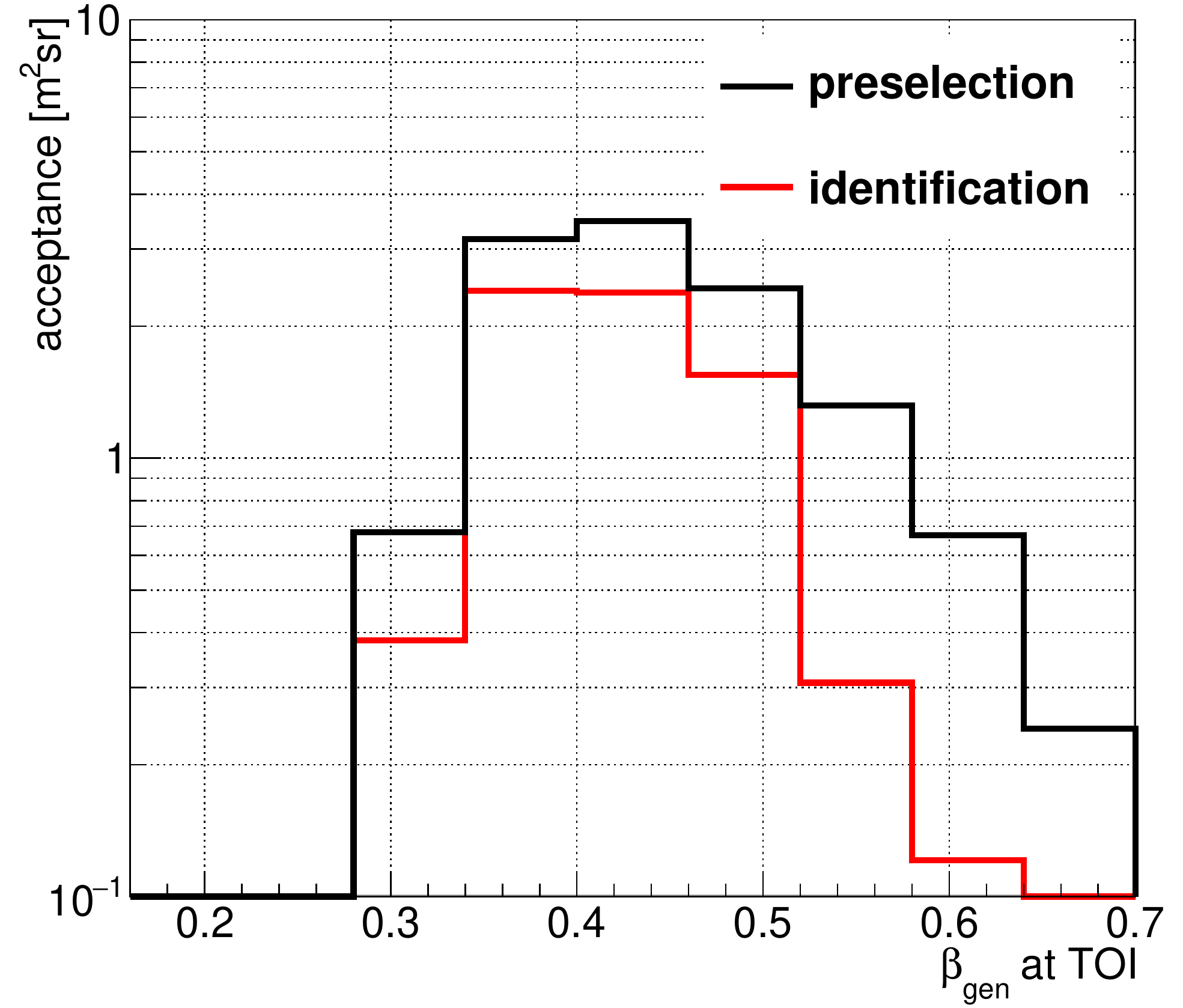}
\caption{\label{f-acceptance}The GAPS acceptance as a function of $\beta_{\text{TOI,gen}}$ for antihelium-3 nuclei after preselection and identification cuts.}
\end{figure}

\section{Sensitivity calculation\label{s-sens}}

\subsection{Atmospheric simulations}
\label{sec-atmosphere}

To determine the number of background events passing antihelium-3 nuclei selection, the identification acceptance calculation for antihelium-3 nuclei and the various background channels at the TOI need to be combined with the anticipated background flux levels. For this purpose, the TOI background fluxes were determined using a separate {\tt Geant4} simulation, based on {\tt PLANETOCOSMICS}~\cite{planeto}, that propagates geo\-magnetically- and solar-modulated cosmic-ray fluxes~\cite{Strong_2007} from the top-of-the-atmosphere (TOA) to TOI. The background fluxes were simulated for the expected LDB float altitude of 37\,km above Antarctica during December. The antiproton fluxes include a contribution from atmospherically produced antiprotons. This model was validated with available data, including the 2012 pGAPS flight~\cite{von_Doetinchem_2014}, and it determines the energy loss and survival probability of antihelium-3 nuclei traversing the atmosphere as well as the energy and angular distributions of background particles. 

Combining the TOI background fluxes with measurement time and their corresponding acceptances to pass all antihelium-3 nucleus selection criteria determines the required background rejection level. The $-\ln(L)$ selection criterion for each angular range was chosen such that one detected antihelium-3 nucleus provides an unambiguous discovery. Fig.~\ref{f-acceptance} compares the acceptance for antihelium-3 nuclei after preselection cuts with the acceptance after all identification cuts. The antihelium-3 nuclei identification efficiency is on the level of about 50\% for the peak region around $\beta_{\text{TOI,gen}}\approx0.34-0.52$. To estimate the number of background events passing cuts, the background acceptances after all identification cuts are integrated with the TOI background fluxes. A detailed publication on the atmospheric studies is forthcoming. 

\subsection{Sensitivity estimate}
\label{sec-sensitivity}

\begin{figure}
\centering
\includegraphics[width=\textwidth]{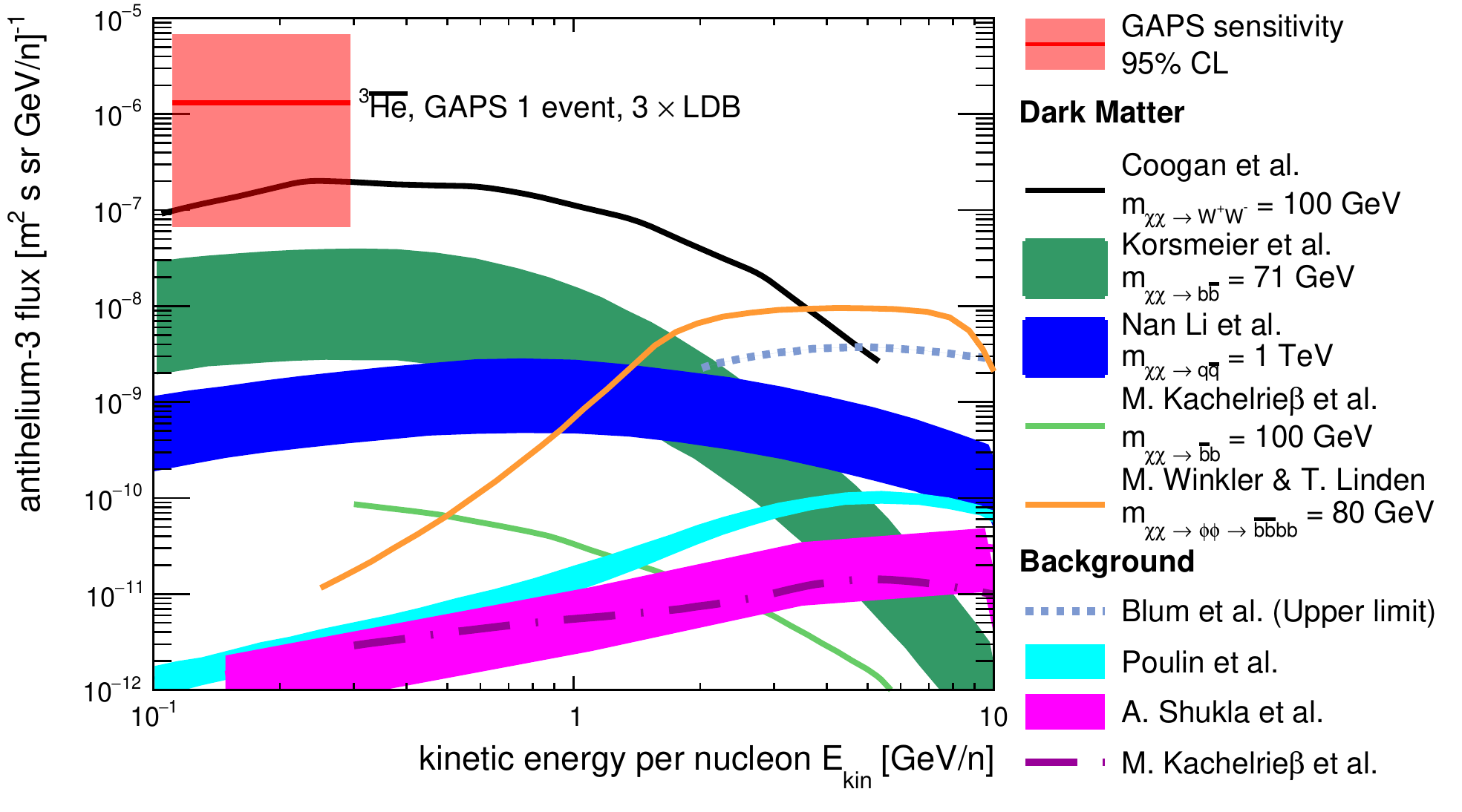}
\caption{\label{f-sensitivity}The solid red line shows the single event sensitivity of GAPS to antihelium-3 nuclei (95\% confidence level) for three LDB flights of 35\,days each. The red box indicates the upper and lower bounds of the 95\% confidence level. Also shown are the antihelium-3 flux predicted by a variety of dark matter~\cite{Korsmeier_2018,Coogan_2017,Blum_2017,Ding_2019,winkler2020dark} and standard astrophysical background~\cite{2018arXiv180808961P, Kachelriess:2020uoh, Shukla_2020} models. For theoretical predictions, the error bands illustrate uncertainties in the coalescence momentum, but also include propagation uncertainties.}
\end{figure}

The number of predicted mean background events (spurious events that pass the antihelium-3 nuclei cuts) after trigger, preselection, and identification cuts is on the order of about $10^{-3}$ for one LDB flight of 35\,days. The subsequent estimation of sensitivity was done with a Bayesian analysis~\cite{Feldman_1998}. Combining the number of background events $b$ with the expectation of one detected antihelium-3-nucleus-like event ($n=1$), the antihelium-3 flux sensitivity $S$ of the GAPS experiment can be calculated for a given confidence interval. The mean value of $S$ is calculated from: 
\begin{equation}
S=\frac{n-b}{\bar A_{\text{id}} T \Delta{E} \epsilon_{\text{geo}} \epsilon_{s}}.
\end{equation}
Here, $T$ is the observation time (three 35-day LDB flights = 105\,days). $\bar A_{\text{id}}$ is the average antihelium-3 nuclei identification acceptance in the TOA kinetic energy range of 0.11--0.3\,GeV/$n$. To determine the GAPS acceptance in this TOA energy range, the antihelium-3 nuclei identification acceptance as a function of $\beta_{\text{TOI,gen}}$ was mapped to the TOA kinetic energy per nucleon using the results of the atmospheric studies described in Sec.~\ref{sec-atmosphere}. $\epsilon_{\text{geo}}$ is the geo\-magnetic cutoff efficiency for antihelium-3 nuclei ($\approx$0.85 in the TOA energy range)~\cite{PvD_Geomagnetic}. $\epsilon_{s}$ is the atmospheric survival probability for antihelium-3 nuclei, which describes the probability of an antihelium-3 nucleus to traverse the atmosphere without being absorbed ($\approx$0.5 averaged across the TOA energy range). The corresponding antihelium-3 nuclei single event sensitivity is $1.3^{+4.5}_{-1.2}\cdot 10^{-6}$\,m\textsuperscript{-2}sr\textsuperscript{-1}s\textsuperscript{-1}(GeV/$n$)\textsuperscript{-1} (95\% confidence level). For one 35-day LDB flight, the projected GAPS antihelium-3 nuclei sensitivity is $4.0^{+13.3}_{-3.8}\cdot 10^{-6}$\,m\textsuperscript{-2}sr\textsuperscript{-1}s\textsuperscript{-1}(GeV/$n$)\textsuperscript{-1} (95\% confidence level). The uncertainties in the projected sensitivities are estimated using the upper and lower limits of true antihelium-3 nuclei detections from the 95\% confidence interval, based on the calculated mean number of background events. Fig.~\ref{f-sensitivity} shows the three-flight sensitivity in comparison with antihelium-3 fluxes predicted by a variety of dark matter~\cite{Korsmeier_2018,Coogan_2017,Blum_2017,Ding_2019,winkler2020dark} and astrophysical background~\cite{2018arXiv180808961P, Kachelriess:2020uoh, Shukla_2020} models. Within the 95\% confidence interval, three GAPS flights have the potential to discover dark matter models annihilating into $W^+W^-$~\cite{Coogan_2017}.

\subsection{Future work}
The current identification technique does not exploit the rejection power associated with exotic-atom de-excitation X-rays, which are an important component of the GAPS antideuteron detection concept~\cite{AramakiSensitivity}. Recently the exotic-atom cascade model~\cite{Aramaki2013} was extended to include antihelium-3 nuclei. This model indicates high yields (about 97\%) for the relevant antihelium-3 nuclei X-rays (43.5, 63.5\,keV). Efforts are currently underway to improve the rejection power by exploiting this X-ray signature of antihelium-3 nuclei stopping in the GAPS tracker.

Studies are also planned to determine the sensitivity of GAPS to cosmic antihelium-4 nuclei. Due to the higher secondary multiplicity in antihelium-4 nuclei events, the event variables used to identify antihelium-3 nuclei are expected to provide even stronger rejection of background particles when applied to antihelium-4 nuclei.

\section{Conclusion and outlook\label{s-conc}}

Low-energy cosmic antihelium nuclei provide an ultra-low background signature of dark matter. Based on full instrument simulation, event reconstruction, and realistic atmospheric influence simulations, a projected GAPS flux sensitivity to antihelium-3 nuclei, assuming the detection of one event in three 35-day LDB flights, was determined to be $1.3^{+4.5}_{-1.2}\cdot 10^{-6}$\,m\textsuperscript{-2}sr\textsuperscript{-1}s\textsuperscript{-1}(GeV/$n$)\textsuperscript{-1} (95\% confidence level) in the energy range of 0.11--0.3\,GeV/$n$. The GAPS sensitivity to antihelium-3 extends to lower energies than any previous experiment, complementing the exclusion limits set by BESS-Polar and ongoing searches with AMS-02. Due to its orthogonal systematic uncertainties and sensitivity to the lower-energy range, where the predicted contribution from new-physics models is highest, GAPS will provide crucial input to interpret the AMS-02 candidate events. This unique sensitivity can be further enhanced by increasing the tracker active area (instrumenting with more Si(Li) detectors), increasing flight times, and improving the background suppression techniques. Future GAPS missions, such as through the NASA Pioneer program, would allow expanding this sensitivity to the $\mathcal{O}(10^{-7})$\,m\textsuperscript{-2}sr\textsuperscript{-1}s\textsuperscript{-1}(GeV/$n$)\textsuperscript{-1} flux range.

\section{Acknowledgments}

This work is supported in the U.S. by NASA APRA grants (NNX17AB44G, NNX17AB45G, NNX17AB46G, and NNX17AB47G) and in Japan by JAXA/ISAS Small Science Program FY2017.
P. von Doetinchem received support from the National Science Foundation under award PHY-1551980.
H. Fuke is supported by JSPS KAKENHI grants (JP17H01136 and JP19H05198) and Mitsubishi Foundation Research Grant 2019-10038. K. Perez and M. Xiao are supported by Heising-Simons award 2018-0766.
F. Rogers is supported through the National Science Foundation Graduate Research Fellowship under Grant No. 1122374.
Y. Shimizu receives support from JSPS KAKENHI grant JP20K04002 and Sumitomo Foundation Grant No. 180322.
This work is supported in Italy by Istituto Nazionale di Fisica Nucleare (INFN) and by the Italian Space Agency through the ASI INFN agreement no. 2018-28-HH.0: ``Partecipazione italiana al GAPS - General AntiParticle Spectrometer".
The technical support and advanced computing resources from the University of Hawaii Information Technology Services -- Cyberinfrastructure are gratefully acknowledged.

\section*{References}

\footnotesize
\bibliography{referencesetal}

\end{document}